# Nonlinear optical response spatial self-hase modulation in MoTe$_2$: correlations between $\chi^{(3)}$ and mobility or effective mass

Lili Hu[1], Fei Sun[1,2], Hui Zhao[1,2], Jimin Zhao[1,2,3]*

[1] *Beijing National Laboratory for Condensed Matter Physics, Institute of Physics, Chinese Academy of Sciences, Beijing 100190, China*

[2] *School of Physical Sciences, University of Chinese Academy of Sciences, Beijing 100049, China*

[3] *Songshan Lake Materials Laboratory, Dongguan, Guangdong 523808, China*

* Corresponding author: jmzhao@iphy.ac.cn

## Abstract

We report an unambiguous observation of third-order nonlinear optical effect, spatial self-phase modulation (SSPM), in a MoTe$_2$ dispersion. The values of the third order nonlinear optical coefficients effectively for one layer MoTe$_2$, $\chi^{(3)}_{\text{one-layer}}$, are obtained through the SSPM method at excitation wavelengths 473, 532, 750 and 801 nm, respectively. The wind-chime model is used to explain the ring formation time. The wavelength dependence of $\chi^{(3)}_{\text{one-layer}}$ compares well with the photo-absorption spectra. Significantly, we find a correlation between $\chi^{(3)}$ and the carrier mobility $\mu$ or effective mass $m^*$, which again further supports the laser-induced ac electron coherence in 2D materials.

Transition metal dichalcogenides (TMDs), denoted as MX$_2$, such as MoS$_2$, WS$_2$, MoSe$_2$, WSe$_2$ and MoTe$_2$ are a group of 2D materials with gapped electronic band structures and high carrier mobilities [1-4]. They draw great interests in optics and optoelectronics investigations, which calls for comprehensive understanding of the nonlinear optical responses of 2D carriers. So far, there is no report on the SSPM of MoTe$_2$, which hinders its applications in nonlinear optics.

The third-order nonlinear optical coefficients $\chi^{(3)}$ of quantum materials can be measured by only a few experimental methods, including spatial self-phase modulation (SSPM) [5-11], third-harmonic generation (THG) [12,13], Z-scan [14], and four-wave mixing (FWM) [15]. SSPM is a third-order nonlinear optical effect [5] found previously in liquid crystal, dye solution [8] and dispersions of 2D materials [5], where the refractive index $n$ of the medium is linearly dependent on the laser intensity $I$ as $n = n_0 + n_2 \times I$. By SSPM method, $\chi^{(3)}$ can be obtained conveniently through measuring the laser intensity dependence of SSPM ring numbers $N$ as shown in Eq. (1). Compared with other coherent methods like THG and FWM, the third-order optical qualities in SSPM are strong and have excellent contrast ratio, making SSPM an effective tool to characterize the third order nonlinear optical property of materials. Recently SSPM effects in a few 2D material dispersions have been investigated, including graphene [6], $MoS_2$ [5], $MoSe_2$ [7], black phosphorous [10], $Bi_2Se_3$ [16] and graphite [11]. In the recent investigation of SSPM for $MoS_2$, it was discovered the SSPM effect is closely related to the confined behavior of 2D electrons [5]. Here, we report the SSPM effect in a 2H-$MoTe_2$ dispersion to unveil the optical and electronic properties of $MoTe_2$. The third order nonlinear coefficients $\chi^{(3)}$ are obtained at wavelengths 473 nm, 532 nm, 750 nm and 801 nm. The formation dynamics of the SSPM pattern are also investigated.

In the experiment, micrometer-sized 2H-$MoTe_2$ powders (LSKYD, www.kydmaterials.com/en/index.php) are dispersed in NMP (N-methyl-2-pyrrolidone) solvent with a concentration of 0.1 g/L. The X-ray diffraction (XRD) pattern is shown in Fig. 1. Comparing with the powder diffraction file standard spectrum, we label the XRD peaks, which compare very well with those of the 2H phase of $MoTe_2$. The crystal structure of 2H-$MoTe_2$ is shown in the left inset of Fig. 1. The scanning electron microscope (SEM) image indicates the diameter of the MoTe2 flakes is approximately 2 μm, as shown in the right inset of Fig. 1. After ultrasonic treatment of the $MoTe_2$ dispersion for 1 hour, a 1 cm-thick cuvette is filled with the dispersion solution and then it is used for the SSPM investigation. Each cuvette of suspension is freshly prepared and investigated in the experiment for less than 0.5 hour, which guarantees that the concentration of the suspension is relatively a constant during the measurements. Continuous laser beams at 473, 532, 750 and 801 nm are used, respectively. The experimental setup is schematically shown in Fig. 2(a), with a lens focal length of 20 cm. The distance between the lens and the cuvette is 17.5 cm and the distance between the cuvette and the screen is 3.2 m.

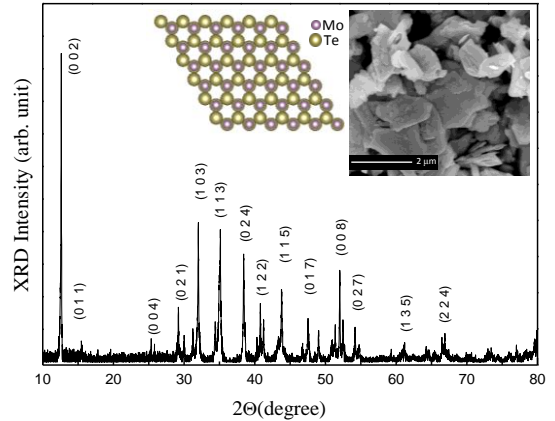

**Fig. 1.** The XRD pattern of our $MoTe_2$ flakes, which verifies a lattice structure with 2H symmetry group. Left inset: schematic illustration of the 2H-$MoTe_2$ lattice structure, which is viewed from the (0, 0, 1) direction. Right inset: The SEM image of $MoTe_2$ flakes before being dispersed in the solvent.

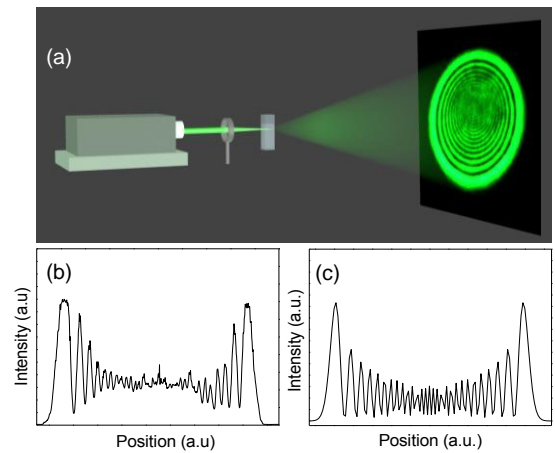

**Fig. 2.** (a) Illustration of our experimental setup for observing SSPM. The SSPM diffraction pattern shown is obtained with an incident continuous wave laser beam of 532 nm wavelength at a power of 150 mw. (b) The (horizontal direction) radial intensity distribution

of the SSPM diffraction pattern shown in (a). (c) Calculated (horizontal) radial intensity distribution for the SSPM pattern.

A typical SSPM pattern is observed using a 532 nm laser beam at 150 mW, which is shown in Fig. 2(a). The intensity distribution along the horizontal direction is plotted in Fig. 2(b). We also calculate the SSPM pattern with parameters according to the experimental conditions. The result is shown in Fig. 2(c). The good agreement between Fig. 2(b) and (c) demonstrates that we are observing the SSPM effect unambiguously.

Similarly, typical patterns generated at four different wavelengths respectively are shown in Fig. 3(a) – (d). We record the ring number N at different laser intensity I for four wavelengths as shown in Fig. 3(e). It can be seen that the ring number N has a linear dependence on the laser intensity I, which also proves what we have observed is the SSPM fringes. From the linear relation between the ring number N and the laser intensity $I$ (Fig 3(e)), the third order nonlinear optical coefficient $\chi^{(3)}$ can be obtained from

$$\chi^{(3)} = \frac{c\lambda n_0}{2.4\times10^4 \pi^2 l} \times \frac{dN}{dI} , \qquad (1)$$

where $n_0 = 1.47$ is the refractive index for NMP solvent, $c$ is light speed in vacuum, $\lambda$ is the laser wavelength, and l is the distance that the laser beam travels in the cuvette. The number of effective MoTe$_2$ layers $M$ that the out-going laser beam has passed is estimated to be 180 [5]. By $\chi^{(3)} = M^2 \cdot \chi^{(3)}_{one-layer}$, the effective $\chi^{(3)}_{one-layer}$ for one-layer MoTe$_2$ is measured to be $1.88 \times 10^{-9}$ e.s.u. at 473 nm, $1.30 \times 10^{-9}$ e.s.u. at 532 nm, $1.14 \times 10^{-9}$ e.s.u. at 750 nm and $0.98 \times 10^{-9}$ e.s.u. at 801 nm, which is shown in Fig. 3 (f).

As shown in Fig. 3 (e), the slopes of the fitting lines decrease gradually with the increasing wavelengths from 473 nm to 801 nm, whose tendency agrees well with the wavelength dependent photo-absorption spectra as shown in Fig. 3 (f) (adapted from Ref. 17). In the SSPM investigation previously reported for MoS$_2$ dispersions [5], the slopes of the lines start to increase prominently when the incident photon energy exceeds the bandgap, demonstrating the SSPM effect is directly correlated with the electron excitations in the 2D material sheets [5]. Here the photon energies of the four wavelengths are larger than the bandgap of MoTe$_2$ (1.0 eV) [17]. Hence the laser beams of the above four wavelengths can be absorbed prominently in MoTe$_2$ flakes dispersed, whereby each photon generates a photocarrier. As a result, we observe prominent SSPM effect for all the four wavelengths. Such SSPM effect exhibits similar wavelength dependence as that of the photo-absorption. Such behavior demonstrates the involvement of electrons in the SSPM effect and echoes that of the electronic coherence in the SSPM of MoS$_2$ [5]. In Ref. [5], the polarization dependence control experiment has demonstrated an evidence of laser beam driven flake rotation [5].

Next we demonstrate the emergence processes of the SSPM ring patterns. The incident laser beam power is 115 mW. The time evolution of the ring patterns is shown in Fig. 4(a). The ring formation dynamics are fitted by the exponential decaying model as

$$N = Ae^{-t/\tau_{rise}}, \qquad (2)$$

where $N$ is the ring number, $\tau_{rise}$ is the rise time for ring formations and $A$ is a constant. As shown in Fig. 4 (b), (c) and (d), the $\tau_{rise}$ for 473, 532 and 750 nm are fitted to be 0.13, 0.23, and 0.25 s respectively. In the wind-chime model [5] the ring-formation time T is given as follows,

$$T = \frac{\varepsilon_r \pi \eta \xi R c}{1.72(\varepsilon_r - 1) I h}, \quad (3)$$

where $\varepsilon_r$ is the relative dielectric constant for MoTe$_2$, $\eta = 1.7 \times 10^{-3}$ Pa·s is the viscosity coefficient for NMP solvent, $\xi \sim 0.1$ in our experiment. $R = 1$ μm is the average radius of the MoTe$_2$ disks, and I is the laser intensity ($I_{473} = 181$ W/cm$^2$, $I_{532} = 219$ W/cm$^2$ and $I_{750} = 252$ W/cm$^2$). We estimate the average thickness of a MoTe$_2$ disk to be $h = 0.15$ μm. The experimental value of ring formation time $T$, the minimal time for reaching the largest number of rings [5], is estimated to be 0.45, 0.6, and 0.62 s, respectively, according to the measurement results shown in Fig. 4(b-d). Thus, the value of electrostatic constant $\varepsilon_r$ can be calculated using Eq. (3), yielding a value of $\varepsilon_r = 4.20$, 1.90, and 1.66, respectively. The measured values of ring formation time again support the validity of the "wind-chime" model for the MoTe$_2$ dispersion. In addition, the rise time for ring formation $\tau_{rise}$ is measured directly from the experiment. In Fig. 4 (b-d), the values of $\tau_{rise}$ are 0.13 s at 473 nm, 0.23 s at 532 nm and 0.35 s at 750 nm, respectively, which naturally show similar laser wavelength dependence as $T$. The dependence for both time constants can be explained by the wavelength dependence of $\varepsilon_r(\omega)$ in Eq. (3). The advantage of introducing the exponential data fitting here is that this provides a likely more valid way of determining the $T$ by finding the cross point between the experimental data of observing the largest number of rings with that of the exponential fitting curve. Such a way of data analysis reduces the experimental error in reading out the time T when the largest number of rings are reached.

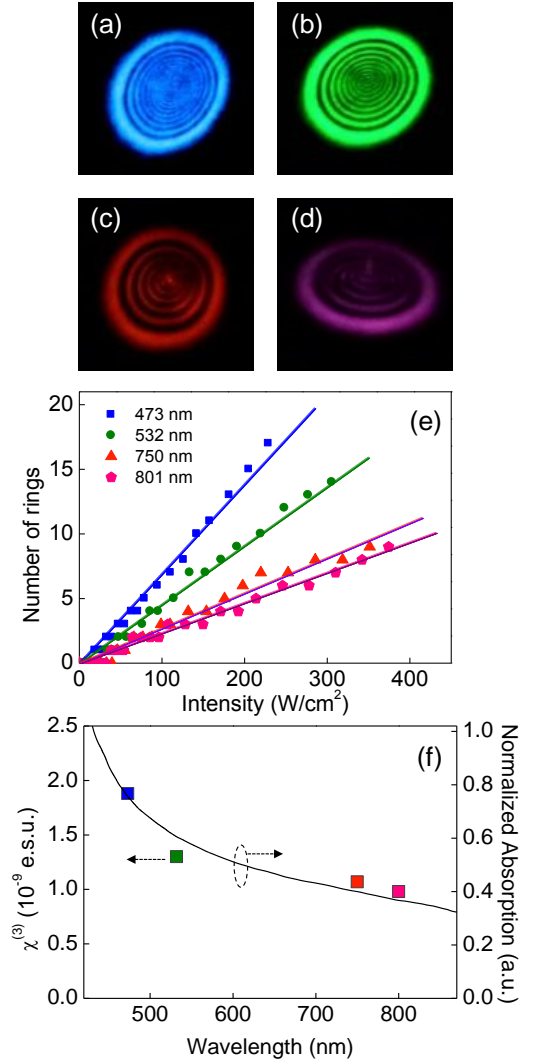

**Fig. 3.** The SSPM ring patterns at (a) 473 nm, (b) 532 nm, (c) 750 nm and (d) 801 nm. (e) The ring number N depending on the laser intensity *I*. The solid lines are fitting lines to the experimental data. (f) The wavelength dependence of $\chi^{(3)}$ (square dots). The solid curve is an absorption spectrum adapted from Ref. 17.

Besides obtaining the third order nonlinear optical coefficient, our results can also provide some related information about the band structure of the materials, by varying the laser wavelengths. Because 2D materials are promising for future application in electronics and photonics, and the monolayer and few-layer 2D materials can be prepared efficiently through exfoliation [18], our finding has a broad application potential in optoelectronics and photonics, including all-optical switching [5].

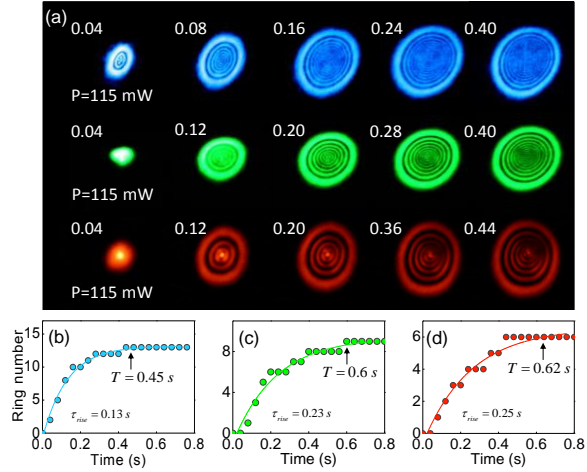

**Fig. 4.** The formation dynamics of SSPM patterns at $P_{473nm}$ = $P_{532nm}$ = $P_{750nm}$ = 115 mW. (a) Time-sliced process of multiple ring emergence. (b-c) Evolution of ring numbers with time at 473, 532 and 750 nm, respectively.

Finally, microscopically, we had contemplated that the electronic coherence is correlated to the carrier mobility [5,6] and effective mass. The carrier mobility characterizes the carrier's motion capability under an external field. It mainly relies on the effective mass and the scattering of the carriers. In the third-order nonlinear optical response SSPM, the ac non-local electronic coherence is established through the driven motion of the carriers in the coherent laser field. We expect that the $\chi^{(3)}$ is positively correlated with the carrier mobility and negatively correlated with the carrier effective mass, because the easier the carriers move following the light field, and the less scattering of the carriers during the motion, the higher the value of $\chi^{(3)}$. Here we summarize the values of $\chi^{(3)}$ obtained by SSPM experiment versus carrier mobility $\mu$ and effective mass $m^*$ in Fig. 5, including values for graphene [6], black phosphorous [10], $MoS_2$ [5], $MoSe_2$ [8] and $MoTe_2$ (this work). The values of $\mu$ and $m^*$ are obtained from Ref. 19-28. Qualitatively, a positive correlation between $\chi^{(3)}$ and $\mu$ and a negative correlation between $\chi^{(3)}$ and $m^*$ are clearly observed, respectively. The red solid line (curve) in Fig. 5(a) and 5(b) are phenomenological fittings to the data. The exact relations are determined by the excited state electronic structure in momentum space of the quantum materials. More thorough quantitative investigation deserves further studies, calling for theoretical derivations and calculations. Note that $\chi^{(3)}$ is an optical parameter, $\mu$ is an electrical property, and $m^*$ is defined by the excited state electronic structure. Thus, if we omit the subtle differences and qualitatively obtain the overall correlations, our results connect a nonlinear optical property to the electronic properties. Hence, our observation adds a new support to the laser induced electron coherence in 2D materials during the SSPM process. By uncovering the correlation between $\chi^{(3)}$ and the electronic properties, our findings also help to predict and discover new materials with exceptional nonlinear optical properties.

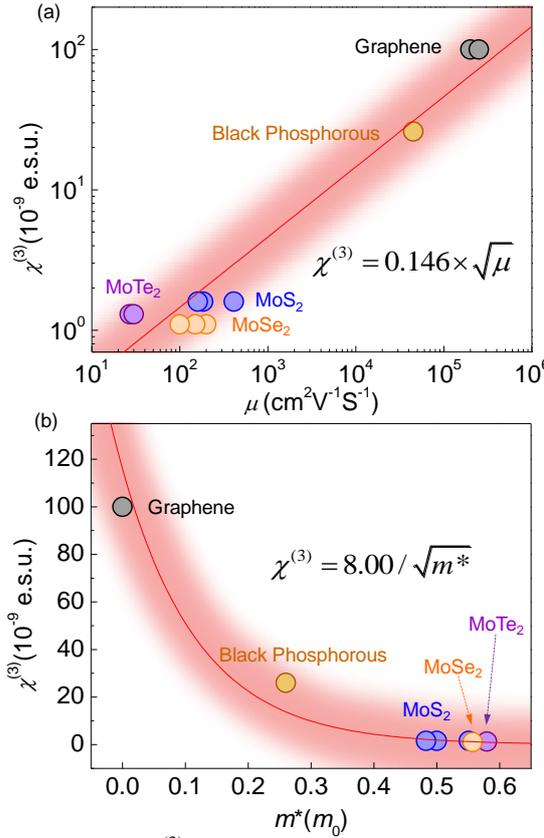

**Fig. 5.** The $\chi^{(3)}$ versus carrier mobility μ (a) and effective mass m$^*$ (b) of several 2D layered materials. Except for the black phosphorous [10], all the values of the $\chi^{(3)}$ are measured by our group [5-7], which have systematic consistency in measurements. The value of $\mu$ and $m^*$ are obtained from Ref. 19-28. The red line and curve are phenomenological fittings to the data.

In summary, we investigate the SSPM effects of MoTe$_2$ dispersion solution at four wavelengths 473, 532, 750, and 801 nm above the bandgap of MoTe$_2$. By SSPM method, we obtain the $\chi^{(3)}$$_{\text{one-layer}}$ of MoTe$_2$ at the above four wavelengths. The ring formation time in MoTe$_2$ dispersions confirms the universal applicability of wind chime model in explaining the SSPM in 2D materials dispersions. Also our investigation shows the SSPM is a general method to investigate the nonlinear optical properties of 2D materials. Significantly, we observe a qualitative correlation between $\chi^{(3)}$ and $\mu$, or between $\chi^{(3)}$ and $m^*$, which sheds new light on the quantitative understanding of the laser-induced ac non-local electron coherence in layered 2D materials.

**Funding.** National Key Research and Development Program of China (2016YFA0300303, 2017YFA0303603); Beijing Natural Science Foundation (4191003); National Natural Science Foundation of China (11574383, 11774408); Strategic Priority Research Program of CAS (XDB30000000); External Cooperation Program of Chinese Academy of Sciences (GJHZ1826); CAS Interdisciplinary Innovation Team.